# Ultrafast quasiparticle dynamics in superconducting iron pnictide CaFe$_{1.89}$Co$_{0.11}$As$_2$

Sunil Kumar,[1,§] L. Harnagea,[2] S. Wurmehl,[2] B. Buchner,[2] and A. K. Sood[1,*]

[1]*Department of Physics, Indian Institute of Science, Bangalore 560012, India*
[2]*Leibniz-Institute for Solid State and materials Research, Dresden, D-01171 Dresden, Germany*

Nonequilibrium quasiparticle relaxation dynamics is reported in superconducting CaFe$_{1.89}$Co$_{0.11}$As$_2$ single crystal using femtosecond time-resolved pump-probe spectroscopy. The quasiparticle dynamics reflects a three-channel decay of laser deposited energy with characteristic time scales varying from few hundreds of femtoseconds to order of few nanoseconds where the amplitudes and time-constants of the individual electronic relaxation components show significant changes in the vicinity of the spin density wave (T$_{SDW}$ ~ 85 K) and superconducting (T$_{SC}$ ~ 20 K) phase transition temperatures. The quasiparticles dynamics in the superconducting state reveals a charge gap with reduced gap value of 2Δ$_0$/k$_B$T$_{SC}$ ~ 1.8. We have determined the electron-phonon coupling constant λ to be ~ 0.14 from the temperature dependent relaxation time in the normal state, a value close to those reported for other types of pnictides. From the peculiar temperature-dependence of the quasiparticle dynamics in the intermediate temperature region between the superconducting and spin density wave phase transitions, we infer a temperature scale where the charge gap associated with the spin ordered phase is maximum and closes on either side while approaching the two phase transition temperatures.

## 1. Introduction

High-temperature superconductivity in FeAs-based pnictides [1-3] has attracted immense interest recently. The quasiparticle pairing in these compounds in comparison with the cuprates is being investigated extensively. Various experimental studies have revealed multiple charge gaps in the superconducting state [4-10]. In the last couple of years, femtosecond time-resolved studies on iron pnictides, mostly on Sm-1111 [11,12] and Ba-122 [13-18] type pnictides have provided valuable information on the electronic relaxation dynamics, the multi-band structure and electron-phonon coupling. The common result among these studies is the observation of two electronic relaxation components with decay times from sub-ps to few ps in the spin density wave (SDW) and the superconducting (SC) states whose temperature and fluence-dependences were partly described using the phonon-bottleneck Routhwarf-Taylor (R-T) model [19]. In few experiments [13,15], a much slower third component with its amplitude decreasing at higher temperatures but the decay-time (~ns) largely independent of the temperature and the fluence was also reported, tentatively attributed to spin-lattice interaction [13]. A high-temperature pseudo-phase with a BCS-type pseudo-gap [11,12] and a normal-state-order by suppression of SDW order just before the emergence of superconductivity in underdoped Ba-122 have been argued to exist as a precursor to superconductivity [13].

The undoped Ca-122 system shows an interesting difference in the temperature dependence of resistivity, namely a steep increase at T$_{SDW}$ rather than a decrease in Ba-122 and other types of pnictides [20]. However, this difference is not present in the corresponding doped systems [21,22]. The Ca-122 systems have not been explored hitherto using time-resolved optical spectroscopy. We present results from femtosecond transient reflectivity measurements on electron-doped iron pnictide CaFe$_{1.89}$Co$_{0.11}$As$_2$ and discuss the temperature and fluence dependence of the quasiparticle dynamics in all three phases: the high-temperature normal metal state (T > 85 K), the SDW state (85 K to 20 K) and the superconducting state below 20 K. Three-component electronic relaxation with sub-ps, a few ps and hundreds of ps relaxation times is observed at all temperatures with large variations in their amplitudes and decay times at the superconducting and SDW phase transition temperatures as well as at very low temperatures in the SC phase. These have been discussed in the framework of bi-particle recombination kinetics across a temperature-dependent charge gap Δ(T) of the R-T model deriving a reduced charge gap value of 2Δ$_0$/k$_B$T$_{SC}$ ~ 1.8 in the superconducting state. In the intermediate temperature region between T$_{SC}$ and T$_{SDW}$, from our experimental results for the fast sub-ps component, we infer that the charge gap associated with the SDW ordering closes at both the superconducting and SDW phase transition temperatures, a behavior similar to that observed in underdoped Ba-122 system [13]. In the high temperature normal metallic phase of the sample, linear temperature dependence of the fastest relaxation time constant has been used to estimate the electron-optical phonon coupling constant to be ~0.14. Moreover, the amplitudes of the electronic relaxation components show a peak feature in the normal state at ~170 K indicating presence of a pseudo state with characteristic temperature scale of that order in the system.

## 2. Experimental details

Single crystals of $CaFe_{1.89}Co_{0.11}As_2$ used in the present study were grown by high temperature solution growth technique using Sn flux and cleaved into platelet samples with thickness ~0.5 mm and c-axis perpendicular to the surface. A complete characterization for the crystal has been reported earlier [21]. The superconducting and SDW phase transitions occur at temperatures $T_{SC}$ ~ 20 K and $T_{SDW}$ ~ 85 K. The crystal shows structural transition from high temperature tetragonal to low temperature orthorhombic symmetry at $T_S$ ~ 88 K. Degenerate pump-probe experiments in a noncollinear geometry were carried out at 790 nm using 45 fs laser pulses taken from a 1-kHz repetition-rate regenerative amplifier. We have used orthogonally-polarized unfocused pump and probe beams with large beam diameters of ~1.9 mm covering about 80% of the sample surface area. Transient differential reflectivity signals were systematically recorded as a function of time-delay between the pump and the probe pulses at various pump-fluences varying from ~25 $\mu J/cm^2$ to 220 $\mu J/cm^2$ and temperatures from 3.1 K to 300 K with the sample mounted inside a continuous flow liquid-helium optical cryostat. During the entire set of experiments presented here, the pump-probe beam polarizations and the angle of incidence as well as the orientation of the crystal inside the cryostat were not changed.

## 3. Results and discussion

Experimental time-resolved differential reflectivity data at various sample-temperatures taken at a moderate value of the pump-fluence of ~80 $\mu J/cm^2$ are presented in Fig. 1 where the temperature dependence of various exponentially decaying and oscillatory components can be clearly seen at faster time-scale (left panel) and longer time-scale (right panel). Overall, the transients can be decomposed into an electronic part (exponentially decaying) and a coherent phononic (damped oscillatory) part. The fitting procedure is shown in Fig. 2(a) using the following function convoluted with the Gaussian laser pulse,

$$\frac{\Delta R}{R} = \left(1 - e^{-t/\tau_r}\right)\left[\sum_k A_k e^{-t/\tau_k} + \sum_l B_l e^{-t/\tau^p_l} \cos(\omega_l t + \phi_l)\right] \quad (1)$$

The term inside the small brackets takes care of the initial rise of the signals with a rise time of $\tau_r$ ~ 80 fs, found to be the same for all the measurements. The data can be consistently fitted (solid lines in Fig. 1 are fits) with three electronic components ($k$ = 1,2,3) with amplitude $A_k$ and time-constant $\tau_k$, and a combination of one fast (sub-ps time-period) oscillatory component ($l$ = 1) attributed to coherent optical phonon mode and two slow (~100 ps time-period) oscillatory components ($l$ = 2,3) attributed to longitudinal and transverse acoustic phonon modes. Our observation of the longitudinal acoustic phonon mode in Ca-122 systems has allowed us to self-consistently derive the optical penetration depth $\zeta$ at 790 nm as a function of temperature [23]. In the sample studied here the optical penetration depth increases from ~40 nm at the room temperature to ~200 nm at 3.1 K (inset of Fig. 2). We may note that the room temperature value of the penetration depth in the Ca-122 systems is within the range of 25 to 60 nm as reported for other 122-type iron pnictides [17,18,24].

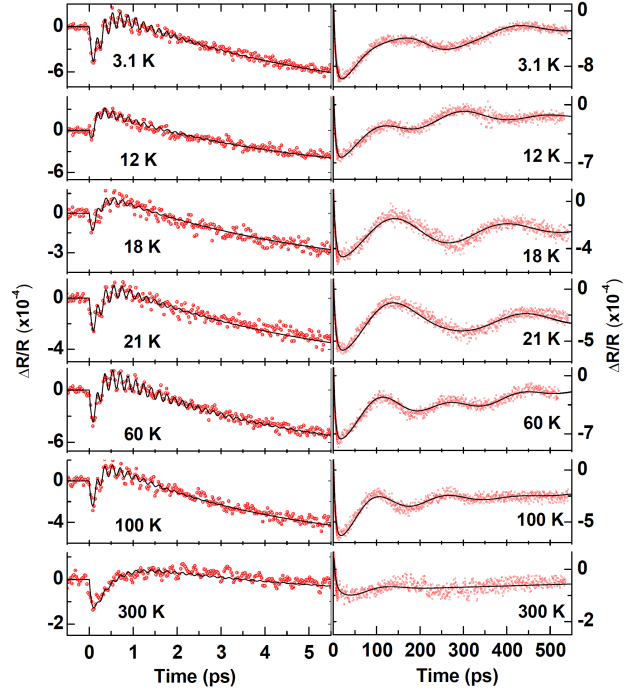

**Fig.1.** (Color online) Transient differential reflection spectra of $CaFe_{1.89}Co_{0.11}As_2$ at various sample temperatures taken using pump-fluence of 80 $\mu J/cm^{-2}$. Solid lines are fits using Eq. 1.

In the present study we have focused on the temperature and fluence dependent electron relaxation dynamics. The results for the temperature-dependence are presented in Fig. 3 where large changes in the values of the parameters can be clearly seen around $T_{SC}$ and $T_{SDW}$. Any isotropic BCS-type charge gap in either superconducting or SDW state is expected to affect the slowest electronic relaxation process related to the smallest band gap in the electronic density of states. However, we find that the dynamics of all the three components (amplitudes $A_{1,2,3}$ and the corresponding relaxation times $\tau_{1,2,3}$) in the superconducting region (T < $T_{SC}$) can be simultaneously described by strong bottleneck R-T model [13,25,26] using a temperature-dependent band gap with reduced gap $\Delta(T)$. Such a feature of the quasiparticle dynamics is unusual and indicative of anisotropic gap and/or multiple gaps in the superconducting state. Theoretically, from band structure calculations, it is known that the Fermi surface of iron pnictides consists of two electron-like pockets at the M (or X)-point and three hole-like pockets at the Γ-point in the Brillouin zone [8]. In the SDW and superconducting states, opening of a gap in these pockets can give rise to the experimental observations.

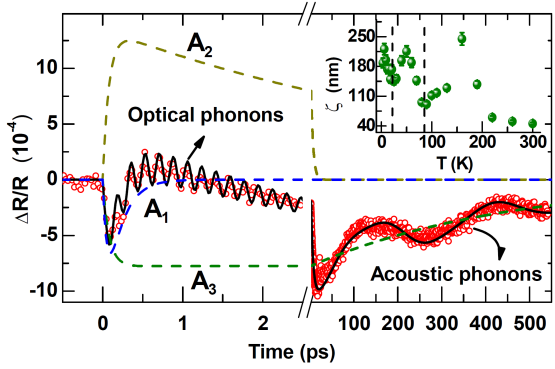

**Fig.2.** (Color online) Fitting procedure of the transient differential reflectivity data (open circles) using Eq. 1 shown for experimental data at 3.1 K. The inset shows the optical penetration depth at 790 nm as a function of temperature.

We may note that the pump-fluence of 80 µJ/cm$^2$ that is used by us to study the temperature dependence of the quasiparticle dynamics in our system is larger than those reported in superconducting Ba-122 systems where the saturation fluence was found to be of the order of few µJ/cm$^2$ to few tens of µJ/cm$^2$ [17,18]. As mentioned before, the optical penetration depth in our sample has been estimated to be quite high ~200 nm at 3.1 K (inset in Fig. 2). In such a case, we can compare the photogenerated quasiparticle density $n_s$ with typical quasiparticle concentration in the sample. At an incident pump-fluence $\Phi$, $n_s$ can be estimated using the relation $n_s = \Phi\lambda_{pr}/hc\zeta$. Using $\Phi = 80$ µJ/cm$^2$, $\zeta = 200$ nm at $\lambda_{pr} = 790$ nm we get $n_s \sim 1.6\times10^{19}$ cm$^{-3}$. Taking unit cell volume in Ca-122 systems to be 170 Å$^3$ [27], we get $n_s \sim 2.7\times10^{-3}$ /unit cell. Typical quasiparticle density $n_0$ in the sample can be calculated using $n_0 = 2N(0)\Delta$ where $N(0)$ is the density of states at the Fermi energy [28]. Taking the smaller gap $\Delta = \Delta_{SC} = 1.8 k_B T_{SC} \sim 3$ meV (our present result) and $N(0) \sim 5$ /eV/unit cell [29] we get $n_0 \sim 3\times10^{-2}$ /unit cell which is almost 10 times larger than $n_s$. Therefore, we are of the view that the fluence of 80 µJ/cm$^2$ can be considered in the weak perturbation range for the R-T model [19,25,26] which we have used in the following subsections to analyze our results at low temperatures. Indeed, the strong non-monotonic temperature dependence of the relaxation parameters below $T_{SDW}$ (Fig. 3) is a clear indication that the thermally generated quasiparticle density plays the major role in our experimental observations.

### 3.1. Quasiparticle dynamics below $T_{SC}$

In the superconducting state ($T < T_{SC}$), the amplitudes $A_k$ of all the three electronic components are minimum at $T_{SC}$ while the decay times $\tau_k$ show a diverging trend at $T_{SC}$ and increase towards lower temperatures (Fig. 3). Such a behavior of the amplitudes and decay times is expected from quasiparticle recombination dynamics across a temperature-dependent band-gap $\Delta(T)$ within the phonon-bottleneck description of the R-T model [25,26]. The behavior of $\tau_1$ and $\tau_2$ in our case is found to be similar to that observed for the fast sub-ps component in the SDW state and slow-picosecond component in the SC state of Ba-122 system [13]. In the following analysis we have used strong bottleneck R-T model equations [25,26] to fit the temperature dependence of the amplitudes together with the time-constants in the superconducting state.

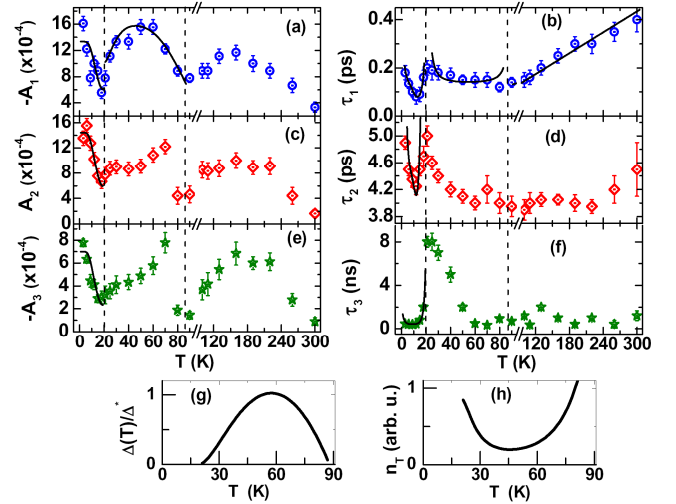

**Fig.3.** Temperature-dependence of amplitudes $A_k$ and decay times $\tau_k$ of three-component electronic relaxation in Eq. 1. The solid lines are theoretical fits as described in the text and the dotted vertical lines mark the nominal superconducting and SDW transition temperatures.

At a constant fluence, the temperature-dependence of quasiparticle dynamics across $\Delta(T)$ is governed by thermally generated quasiparticle density $n_T$, given by [13,25,26]
$$n_T \propto \sqrt{T\Delta(T)} \exp[-\Delta(T)/k_BT] \quad (2)$$
This is related to the experimentally measured amplitude $A(T)$ via the relation
$$n_T \propto [A(T \to 0)/A(T)] - 1 \quad (3)$$
Similarly, the solution of the R-T model in the strong bottleneck regime for the relaxation time $\tau$ is given by [13,25,26]
$$\tau^{-1}(T) = \Gamma(T)\{2n_T + \delta/\varepsilon\, n_T + 1\} \quad (4)$$
where $\delta$ and $\varepsilon$ are temperature-independent fitting parameters and temperature-dependence of $\Gamma$ arises from the temperature-dependence of the phonon decay rate $\gamma$ leading to an expression $\Gamma(T) = \Gamma_0[\Delta(T) + \alpha T\Delta(T)^4]$ (Ref. [13]). Here, the first term results from inelastic scattering of a high energy phonon into a low energy phonon while the second term determines the decay of a high energy phonon with energy $E_p > \Delta$ into two low energy phonons, each with energy less than $\Delta$. The coefficient $\alpha$ has an upper limit of $52/\theta_D^3$ and the second term can be neglected if $\Delta(T) << \theta_D$ [13]. In iron pnictides the Debye temperature $\theta_D \sim 140$ K as estimated from thermal conductivity measurements on Ba-122 [30], which is comparable with $\Delta(T)$ and hence the second term can not be neglected.

Assuming a BCS-like form $\Delta(T) = 2\Delta_0\{1-(T/T_{SC})^2\}$ where $2\Delta_0$ is the zero-temperature value of the superconducting gap, very good fits of the three amplitudes ($A_{1,2,3}$) and corresponding decay times ($\tau_{1,2,3}$) for $T < T_{SC}$ as shown by solid lines in Fig. 3(a-f) with the theory give $2\Delta_0/k_BT_{SC} \sim 1.8$. This value is just half of the BCS limit (the mean field value). Due to strong interband scattering, the involved charge gap should be the smallest one that can be observable in time-resolved studies. In previous time-resolved studies on Sm-1111 [11,12] and Ba-122 [13,15] iron pnictide superconductors, $2\Delta_0/k_BT_{SC} \sim 3$ to 3.7 have been estimated. In a number of recent infrared absorption [4-7] and ARPES [8-10] measurements, multiple superconducting gaps with wide range of $2\Delta_0/k_BT_{SC} \sim 1$ to 10 have been reported. Our estimate of the reduced gap value of 1.8 is similar to that obtained for Ba-122 superconductor studied using ARPES [9] where this low value was inferred to having multi-band superconductivity in iron superconductors with electron pairing between disconnected Fermi surfaces mediated by both orbital and spin fluctuations.

### 3.2. Quasiparticle dynamics in the region $T_{SC} < T < T_{SDW}$

In the region $T_{SC} < T < T_{SDW}$ (Fig. 3) it is noteworthy to note that the amplitudes are highly temperature dependent, particularly the amplitude $A_1$ of the fast component which is decreasing on either side towards $T_{SC}$ and $T_{SDW}$ with a maximum at an intermediate temperature $T^*$. Secondly, we can clearly see an increase in the relaxation times towards $T_{SC}$ as the sample is cooled down in the SDW phase. These are indicative of a temperature dependent charge gap in the SDW state that starts developing at $T_{SDW}$, goes through a maximum in between and then closing at $T_{SC}$ before entering the superconducting phase. Previously, in a study by Chia *et al.*, on underdoped Ba-122 system [13], the fast component (sub-ps relaxation time) related to the SDW order showed its amplitude decreasing at $T_{SC}$ and $T_{SDW}$, and simultaneous quasidivergence in the relaxation time at the two transition temperatures. Such a behavior was associated with suppression of SDW order due to a normal-state-order, a precursor to superconductivity and characterized by temperature scale $T^*$ between $T_{SC}$ and $T_{SDW}$,.

Various experimental studies have provided evidences that the competing antiferromagnetic (AFM) or spin density wave and the superconducting phases in the iron pnictides are macroscopically separated by a first-order transition [31,32]. Fernandes and Schmalian [33] have shown that a superconducting order parameter with symmetry $s^{++}$ does not allow the two phases to coexist around the intersection of the two phases whereas the unconventional $s^{+-}$ state whose gap function changes sign from one Fermi surface sheet to the other, may or may not coexist with AFM depending on the details of the band-structure dispersion relation. Our results in the $T_{SC} < T < T_{SDW}$ region possibly point out the former scenario where the SC and AFM phases are mutually exclusive. Attributing the fast component ($A_1,\tau_1$) in Figs. 3a and 3b to quasiparticle dynamics across a charge gap opening at $T_{SDW}$, we propose a simple model for the temperature-dependent gap that closes at both $T_{SC}$ and $T_{SDW}$ with a maximum in between at a temperature $T^*$:

$$\Delta(T) = \Delta^*\{1-(T/T_{SDW})^2\}\{(T-T_{SC})/(T+T_{SC})\}^2 \quad (5)$$

In Fig. 3g, we have plotted Eq. (5) and the corresponding thermally generated normalized quasiparticle density $n_T$ using Eq. (2) is shown in Fig. 3(h). Substituting Eq. (5) into Eqs. (3) and (4) through Eq. (2), the trends in the temperature-dependence of the amplitude $A_1$ and the time-constant $\tau_1$ in the temperature window between $T_{SDW}$ and $T_{SC}$ could be reproduced (solid lines in Figs. 3a and 3b).

### 3.3 Quasiparticle dynamics in the normal state

In the normal paramagnetic-metallic state ($T > T_{SDW}$), the amplitudes ($A_{1,2,3}$) increase with temperature with a maximum around $\sim 170$ K (Figs. 3a,c,e) whereas, among the three decay times, only the fastest one ($\tau_1$) shows a temperature-dependence (Fig. 3b). The linear temperature-dependence of $\tau_1$ (Fig. 3b) can be used to estimate the electron-phonon coupling constant $\lambda$ using [12] the relation $\tau = \frac{2\pi k_B T}{3\hbar\lambda<\hbar^2\omega^2>}$ where $\lambda<\hbar^2\omega^2>$ is the second moment of the Eliashberg function. Linear fit to $\tau_1$ (solid line in Fig. 3b) gives $\lambda<\hbar^2\omega^2> \sim 57$ meV$^2$, a value about half of that obtained for undoped Sm-1111 pnictide [12] and close to that obtained from Ni-doped and Co-doped superconducting Ba-122 compounds [15,16]. Taking $<\hbar^2\omega^2> \sim 400$ meV$^2$ [12] the electron-phonon coupling constant $\lambda$ is estimated to be $\sim 0.14$, which is too small to explain the superconductivity in iron pnictides within the conventional BCS framework.

The peak feature in the amplitudes $A_{1,2,3}$ at temperature $\sim 170$ K (Fig. 3) can be related to some characteristic temperature scale of this order in the system. One important consequence of the temperature dependent optical penetration depth (inset of Fig. 2) is that the amplitudes of the relaxation components have to be renormalized with respect to it for the same value of energy density absorbed by the sample at various temperatures. By renormalizing the amplitudes $A_{1,2,3}$ in Fig. 3 with respect to $\zeta(T)/\zeta(T = 3.1$ K$)$, we find that below $T_{SDW}$ the behavior of the amplitudes remains the same as before hence does not change the obtained fit parameters in Eqs. 2 to 5. However, the characteristic temperature scale at ~170 K becomes better evident by a sharper peak in the amplitudes.

### 3.4. Fluence-dependent quasiparticle dynamics

The fluence-dependences of the electronic relaxation components ($A_k$,$\tau_k$) for three representative temperatures each from the SC, SDW and normal states are shown in Fig. 4. Solid and dashed lines in Fig. 4 have been drawn as guide to eyes. The slowest relaxation time $\tau_3$ remains nearly fluence-independent in all the three temperature regions (Fig. 4f). In the normal state (T = 160 K) the amplitudes $A_{1,2}$ increase linearly with the fluence while the decay times $\tau_{1,2}$ are fluence-independent. However, in the SC (12 K) and SDW (50 K) states the fluence dependence of the amplitudes $A_{1,2}$ indicates a change in slope at ~130 $\mu J/cm^2$ beyond which they start increasing more rapidly just like in the normal state. Such a behavior can arise due to laser-induced non-thermal destruction of the SC and SDW states [34] without heating the sample above the critical temperatures where the amount of fluence needed for the non-thermal destruction depends on the magnitude of the corresponding order parameter. Also, the excitation energy required to induce such a transition depends on the time scales to reach to quasi-equilibrium between the quasiparticles and the high frequency phonons after the photoexcitation. The order parameter recovers on a sub-ps time scale within which the lattice is not coupled with the electron subsystem and only in the second step of relaxation, (few ps) the electrons adiabatically follow the lattice.

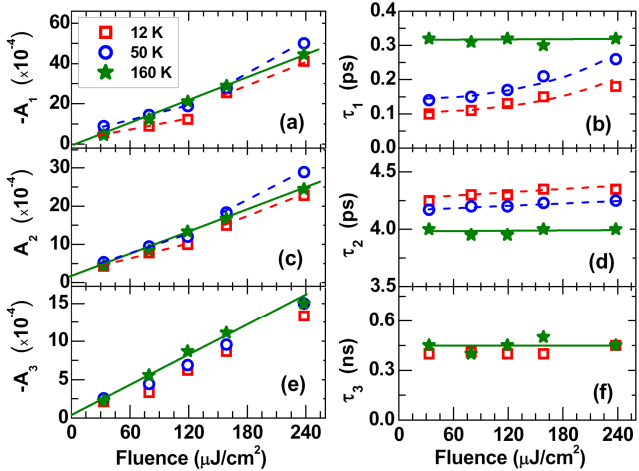

**Fig.4.** Pump-fluence dependence of the amplitudes $A_k$ and decay times $\tau_k$ of fast and slow relaxation components shown for three representative sample temperatures from each region of T < $T_{SC}$, $T_{SC}$ < T < $T_{SDW}$ and T > $T_{SDW}$. Solid and dashed lines are guide to the eyes.

The increase in the fast time constant $\tau_1$ in the SC (12 K) and the SDW (50 K) states with increasing fluence (Fig. 4b) is at odds with the R-T model which predicts that the low-temperature relaxation rate (1/$\tau$) associated with the order parameter recovery increases with the excitation fluence [25,26]. On the other hand, at very high fluences, the increase in the quasiparticle relaxation time ($\tau_1$ in Fig. 4b) can be associated with the nonthermal destruction of the broken symmetry ground states [34], i.e., the system is driven into a normal phase and the corresponding amplitudes $A_1$ and $A_2$ (Figs. 4a and 4c) behave just like in the normal state. Under this situation, the vanishing gap results in an increase in the relaxation time. In the prior pump-probe studies where high repetition rate lasers were used to provide a maximum fluence of ~100 $\mu J/cm^2$, the fluence-dependence of the relaxation times in superconducting Sm-1111 pnictides (Ref. [11,12]) was absent. However, other studies on superconducting Ba-122 pnictides [14,17] where even much smaller fluences were used, the fast relaxation rate was found to increase with increasing fluence, suggesting different relaxation mechanisms in 122 and 1111 type of compounds in the iron pnictide family.

Finally, it is pointed out that in the literature the time-resolved optical studies on 122-type iron pnictides, mostly on the Ba-122 compounds have shown a mix of results for the temperature and fluence dependent quasiparticle dynamics in the superconducting state. In the undoped case, the $BaFe_2As_2$ system shows a single electronic relaxation process that follows the R-T model description for its amplitude and relaxation time as a function of temperature below $T_{SDW}$ [13,24]. Whereas, in the doped Ba-122 systems (electron-doped by Co or Ni doping and hole-doped by K doping), two to three relaxation components having differences in the temperature and fluence dependence of the relaxation parameters from one study to another, have been reported [13-18]. In contrast, our experimental results on the undoped [35] and doped (present study) compound show that in Ca-122 systems, the three-component electronic relaxation dynamics is intact throughout the temperature range of 3 to 300 K with large variations in the amplitudes and corresponding time-constants of the individual relaxation process at $T_{SC}$ and/or $T_{SDW}$ due to opening of charge gaps in the Fermi surface below the respective transition temperatures.

### 4. Conclusions

In this paper, we have investigated the transient differential reflectivity of doped iron pnictide $CaFe_{1.89}Co_{0.11}As_2$ to extract the relaxation dynamics of the quasiparticles in the temperature range of 300 to 3.1 K. The quasiparticle dynamics involves three-component relaxation, where, both the amplitudes and the time-constants of the three components vary significantly at $T_{SC}$ and $T_{SDW}$. In the superconducting state, using strong bottleneck R-T model we have derived a gap value of $2\Delta_0/2k_BT_{SC}$ ~ 1.8, whereas, the linear temperature dependence of the fast time-constant in the normal state has been used to estimate the electron-phonon coupling constant of ~0.14. At intermediate temperatures between $T_{SC}$ and $T_{SDW}$ we infer that the charge gap associated with the spin ordered phase closes while approaching the superconducting and spin density wave phase transition temperatures.

**Acknowledgements -** AKS and SK acknowledge Department of Science and Technology, India for financial assistance.

§*sunilvdabral@gmail.com*
\**asood@physics.iisc.ernet.in*

[1] Y. Kamihara, T. Watanabe, M. Hirano, H. Hosono, J. Am. Chem. Soc. 130, 3296 (2008).
[2] Z.-A. Ren, G.-C. Che, X.-L. Dong, J. Yang, W. Lu, W. Yi, X.-L. Shen, Z.-C. Li, L.-L. Sun, F. Zhou, Z.-X. Zhao, Europhys. Lett. 83, 17002 (2008).
[3] A. J. Drew, F. L. Pratt, T. Lancaster, S. J. Blundell, P. J. Baker, R. H. Liu, G. Wu, X. H. Chen, I. Watanabe, V. K. Malik, A. Dubroka, K.W. Kim, M. Rossle, C. Bernhard, Phys. Rev. Lett. 101, 097010 (2008).
[4] G. Li, W. Z. Hu, J. Dong, Z. Li, P. Zheng, G. F. Chen, J. L. Luo, N. L. Wang, Phys. Rev. Lett. 101, 107004 (2008).
[5] A. Perucchi, L. Baldassarre, S. Lupi, J. Y. Jiang, J. D.Weiss, E. E. Hellstrom, S. Lee, C. W. Bark, C. B. Eom, M. Putti, I. Pallecchi, C. Marini, P. Dore, Eur. Phys. J. B 77, 25 (2010).
[6] T. Fischer, A. V. Pronin, J. Wosnitza, K. Iida, F. Kurth, S. Haindl, L. Schultz, B. Holzapfel, E. Schachinger, Phys. Rev. B 82, 224507 (2010).
[7] E. van Heumen, Y. Huang, S. de Jong, A. B. Kuzmenko, M. S. Golden, D. van der Marel, Europhys. Lett. 90, 37005 (2010).
[8] K. Nakayama, T. Sato, P. Richard, Y.-M. Xu, Y. Sekiba, S. Souma, G. F. Chen, J. L. Luo, N. L. Wang, H. Ding, T. Takahashi, Europhys. Lett. 85, 67002 (2009).
[9] T. Shimojima, F. Sakaguchi, K. Ishizaka, Y. Ishida, T. Kiss, M. Okawa, T. Togashi, C.-T. Chen, S. Watanabe, M. Arita, K. Shimada, H. Namatame, M. Taniguchi, K. Ohgushi, S. Kasahara, T. Terashima, T. Shibauchi, Y. Matsuda, A. Chainani, S. Shin, Science 332, 564 (2011).
[10] D. V. Evtushinsky, D S Inosov, V B Zabolotnyy, M S Viazovska, R Khasanov, A Amato, H-H Klauss, H Luetkens, Ch Niedermayer, G L Sun, V Hinkov, C T Lin, A Varykhalov, A Koitzsch, M Knupfer, B Büchner, A A Kordyuk, S V Borisenko, New J. Phys. 11, 055069 (2009).
[11] T. Mertelj, V. V. Kabanov, C. Gadermaier, N. D. Zhigadlo, S. Katrych, J. Karpinski, D. Mihailovic, Phys. Rev. Lett. 102,117002 (2009).
[12] T. Mertelj, P. Kusar, V. V. Kabanov, L. Stojchevska, N. D. Zhigadlo, S. Katrych, Z. Bukowski, J. Karpinski, S. Weyeneth, D. Mihailovic, Phys. Rev. B 81, 224504 (2010).
[13] E. E. M. Chia, D. Talbayev, J.-X. Zhu, H. Q. Yuan, T. Park, J. D. Thompson, C. Panagopoulos, G. F. Chen, J. L. Luo, N. L. Wang, A. J. Taylor, Phys. Rev. Lett. 104, 027003 (2010).
[14] D. H. Torchinsky, G. F. Chen, J. L. Luo, N. L. Wang, N. Gedik, Phys. Rev. Lett. 105, 027005 (2010).
[15] Y. Gong, W. Lai, T. Nosach, L. J. Li, G. H. Cao, Z. A. Xu, Y. H. Ren, New J. Phys. 12, 123003 (2010).
[16] B. Mansart, D. Boschetto, A. Savoia, F. R.-Albenque, F. Bouquet, E. Papalazarou, A. Forget, D. Colson, A. Rousse, M. Marsi, Phys. Rev. B 82, 024513 (2010).
[17] D. H. Torchinsky, J. W. McIver, D. Hsieh, G. F. Chen, J. L. Luo, N. L. Wang, N. Gedik, Phys. Rev. B 84, 104518 (2011).
[18] L. Stojchevska, T. Mertelj, J.-H. Chu, I. R. Fisher, D. Mihailovic, arXiv:1107.5934 (2012).
[19] A. Rothwarf, B. N. Taylor, Phys. Rev. Lett. 19, 27 (1967).
[20] G. Wu, H. Chen, T. Wu, Y. L. Cie, Y. J. Yan, R. H. Liu, C. F. Wang, J. J. Ing, X. H. Chen, J. Phys.: Condens. Matter 20, 422201 (2008).
[21] L. Harnagea, S. Singh, G. Friemel, N. Leps, D. Bombor, M. A.-Hafiez, A. U. B. Wolter, C. Hess, R. Klingeler, G. Behr, S. Wurmehl, B. Buchner, Phys. Rev. B 83, 094523 (2011).
[22] L. Fang, H. Luo, P. Cheng, Z. Wang, Y. Jia, G. Mu, B. Shen, I. I. Mazin, L. Shan, C. Ren, H.-H. Wen, Phys. Rev. B 80, 140508 (2009).
[23] S. Kumar, L. Harnagea, S. Wurmehl, B. Buchner, A. K. Sood, Europhys. Lett. 100, 57007 (2012).
[24] L. Stojchevska, P. Kusar, T. Mertelj, V. V. Kabanov, X. Lin, G. H. Cao, Z. A. Xu, D. Mihailovic, Phys. Rev. B 82, 012505 (2010).
[25] V. V. Kabanov, J. Demsar, D. Mihailovic, Phys. Rev. Lett. 95, 147002 (2005).
[26] J. Demsar, J. Sarrao, A. J. Taylor, J. Phys.: Condens. Matter 18, R281 (2006).
[27] S. Ran, S. L. Budko, D. K. Pratt, A. Kreyssig, M. G. Kim, M. J. Kramer, D. H. Ryan, W. N. R.-Weetaluktuk, Y. Furukawa, B. Roy, A. I. Goldman, P. C. Canfeld, Phys. Rev. B 83, 144517 (2011).
[28] V. V. Kabanov, J. Demsar, B. Podobnik, D. Mihailovic, Phys. Rev. B 59, 1497 (1999).
[29] E. Z. Kurmaev, J. A. McLeod, A. Buling, N. A. Skorikov, A. Moewes, M. Neumann, M. A. Korotin, Y. A. Izyumov, N. Ni, P. C. Canfield, Phy. Rev. B 80, 054508 (2009).
[30] C. Kant, J. Deisenhofer, A. Günther, F. Schrettle, A. L. M. Rotter, D. Johrendt, Phys. Rev. B 81, 014529 (2010).
[31] H. Luetkens, H.-H. Klauss, M. Kraken, F. J. Litterst, T. Dellmann, R. Klingeler, C. Hess, R. Khasanov, A. Amato, C. Baines, M. Kosmala, O. J. Schumann, M. Braden, J. Hamann-Borrero, N. Leps, A. Kondrat, G. Behr, J.Werner, B. Büchner, Nature Materials 8, 305 (2009).
[32] J. T. Park, D. S. Inosov, C. Niedermayer, G. L. Sun, D. Haug, N. B. Christensen, R. Dinnebier, A. V. Boris, A. J. Drew, L. Schulz, T. Shapoval, U. Wolff, V. Neu, X. P. Yang, C. T. Lin, B. Keimer, V. Hinkov, Phys. Rev. Lett. 102, 117006 (2009).
[33] R. M. Fernandes and J. Schmalian, Phys. Rev. B 82, 014521 (2010).
[34] A. Tomeljak, H. Schafer, D. Stadter, M. Beyer, K. Biljakovic, J. Demsar, Phys. Rev. Lett. 102, 066404 (2009).
[35] S. Kumar, L. Harnagea, S. Wurmehl, B. Buchner, A. K. Sood, (to apear in) J. Phys. Soc. Jpn. (2013).